\documentclass[twocolumn,aps]{revtex4} 
\usepackage{amsmath}
\usepackage{latexsym}
\usepackage{epsfig}

\begin{document}
\title{Confinement from Merons
}
\author{F. Lenz$^{{\rm a}}$,
J. W. Negele$^{{\rm b}}$ and M. Thies$^{{\rm a}}$}
\address{$^{{\rm a}}$ Institute for Theoretical Physics III, 
University of Erlangen-N\"urnberg, 
Staudstrasse 7, 91058 Erlangen, Germany\\ 
$^{{\rm b}}$ Center for Theoretical Physics,
Laboratory for Nuclear Science, and Department of Physics 
Massachusetts Institute of Technology 
Cambridge, Massachusetts 02139, U.S.A.  
}


\date{\today} 
\begin{abstract}
It is shown that an effective theory with meron degrees of freedom produces
confinement in SU(2) Yang Mills theory. This effective theory is compatible
with center symmetry. When the scale is set by the string tension,
the action density and topological susceptibility are  similar  to
those arising
in lattice QCD.

\end{abstract}

\maketitle


\section{Introduction}
Although confinement is one of the most striking and fundamental phenomena
that arises from the deceptively simple QCD Lagrangian, its physical
mechanism has yet to be understood.
One important element is sufficient disorder to drive a large Wilson loop to
produce an area law as seen, for example, in the strong coupling
expansion of lattice QCD, which produces an area law already in lowest order.
However, disorder alone is not the whole story, as evidenced by the fact
that the strong coupling expansion also erroneously yields confinement in U(1)
gauge theory.
A second important feature is center symmetry. By gauge
fixing in continuum QCD up to a residual center symmetry
   or in lattice QCD by multiplying all the links in the time direction on a
single time slice by an element of the center,  it follows that the effective
action has the symmetry $S_{\rm eff}(P) = S_{\rm eff}({\cal Z} P) $ for an
element of the
center,
${\cal Z} $\cite{Yaffe:qf,Svetitsky:1985ye}. A serious candidate for the confinement mechanism should have the potential of realizing this symmetry.
One appealing analytical approach to understanding nonperturbative QCD is
expansion of the path integral for the partition function around stationary
classical solutions and evaluating the fluctuations around these solutions.
Development of highly successful
instanton liquid models and observation of instantons and their zero
modes in lattice QCD  have provided clear insight into how chiral
symmetry breaking arises in QCD \cite{Schafer:1996wv,Negele:1998ev}.
However,  singular gauge instantons fail to produce confinement and, as will be seen below,  also fail to 
produce center
symmetry.

In this work, we reexamine the possibility that merons, another set of
solutions to the classical field equations,  can produce confinement.
As noted long ago \cite{Callan:1978bm}
, merons are
sufficiently disordering that they have the potential to produce confinement.
In contrast to instantons, whose gauge fields fall off as $r^{-3}$ in
singular gauge, merons can only be written in  regular gauge with fields that
decrease as $r^{-1}$.
In the absence of correlations between distant merons, this long-range
gauge field would give rise to an unphysical background  field. Since
analytic treatment of these correlations appears to be intractable,
the behavior of meron ensembles has not previously been analyzed.

\begin{figure}
\epsfig{file=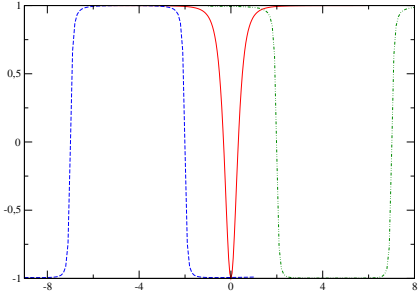, width=.6\linewidth,angle=-90}
\vskip -.3cm
\caption{ Polyakov loop $P(t,0,0)$ as a function of time.  The solid line
corresponds to  one instanton centered  at  the origin.  The dashed and dotted-dashed lines   
show the
results for a meron-meron pair  located on the $t$-axis at $t = -7,-2$ and an antimeron-meron pair at $t=2,7$. }
\label{pol}
\vskip -.3cm
\end{figure}

In SU($N$) gauge theories, the deconfinement transition is associated with a
symmetry property. Center symmetry, a residual
discrete gauge symmetry, is realized in the confined phase  and  broken in
the deconfined phase.  In
SU(2), center symmetry transformations change the sign of the Polyakov loop,
which we define here  along the compact $z$-direction of
Euclidean space,
    $$ P(t,x,y)= \frac{1}{2}\,\mbox{tr}\,P\,\exp\{{\rm i}\oint {\rm d}z A_3(t,x,y,z)\} .$$
   Realization of the center symmetry implies vanishing of the
expectation value of the Polyakov loop  which in turn guarantees an exponential
decrease in the Polyakov loop correlator related to confinement. In the
deconfined phase, the Polyakov loop expectation value is finite. The Polyakov
loop thus serves  as an order parameter. 
In the construction of  ensembles of field configurations  with which
to explore
confinement, it is useful to study the Polyakov loop of the building blocks. As
Fig.  \ref{pol} shows, unlike the winding number, the asymptotic value of the
Polyakov loop of an instanton is not changed in the tunneling process and is
identical for instantons and anti-instantons.  Thus, tunneling processes do not
give rise to a coherent superposition of fields with opposite values of $P$ so
that  a center-symmetric ensemble of field configurations cannot  be
generated in a natural way by superposition of singular gauge instantons. In
contrast, 
the tunneling process  described by a single meron changes the  sign of the
Polyakov loop as can be seen from the figure.  
Furthermore, the asymptotic values of $P$ for a
meron-antimeron pair and a meron-meron pair are of opposite sign.
Thus, merons do not single out either one of the two center-elements of SU(2)
and have the potential of generating a center
symmetric ensemble.

\section{Meron Effective Theory}

In this work, we explore the idea that merons are the essential degrees of
freedom in SU(2) Yang-Mills gauge theory by writing the partition function as a
path integral of an effective action depending on the positions and color
orientations of an ensemble of merons,
\begin{equation}
   \label{pathintegral}
Z = \int {\rm d} z_i {\rm d} h_i e^{- \frac{1}{g^2} S[A(z_i,h_i)]         }\ .
\end{equation}
The gauge field for a meron in Lorentz gauge with its center  at the
origin, after appropriate choice of the coordinate system in color space  and
after regularization of  the singularity,  is given by
\begin{equation}
    \label{mer}
     a_{\mu}(x) = \frac{\eta _{a\mu \nu } x_{\nu}}{x^{2} + \rho ^{2}}
\frac{\sigma^a}{2}.
\end{equation}
Color and space-time dependence are correlated via the 't Hooft tensor $\eta _{a\mu \nu }$ \cite{Schafer:1996wv}.
For vanishing meron size $\rho$, $a(x)$ is a solution of the  Euclidean
classical  field equations \cite{deAlfaro:1976qz}. Antimerons differ in sign if
one of the space-time indices is 0. The action density of merons and
antimerons is given by
\begin{equation}
    \label{acdme}
    s(x) = \frac{1}{2}\, \mbox{tr}\, F_{\mu\nu}  F_{\mu\nu}= 
\frac{3}{2(x^2+\rho^2)^4}\big[x^4+4\,x^2 \rho^2  +8\rho^4\big].
\end{equation}
Unlike instantons, the field-strength decays asymptotically as $1/x^2$,
giving rise to an infrared logarithmic singularity in the action.
For vanishing
meron size,  the action is also logarithmically divergent  in the
ultraviolet, so
we will use $\rho$ as an ultraviolet regulator.  The topological charge density
of a meron or  antimeron
\begin{equation}
    \label{tocdme}
    \tilde s(x) = \pm \frac{1}{2}\, \mbox{tr}\, F_{\mu\nu}\tilde{F}_{\mu\nu}=
\pm \frac{6\rho^2}{(x^2+\rho^2)^4}\big[x^2 + 2\rho^2\big]
\end{equation}
leads to a finite, size-independent topological charge
\begin{equation}
    \label{topc}\nu =  \frac{1}{8\pi^2}\int {\rm d}^4 x \, \tilde{s}(x) = \pm\frac{1}{2} .\end{equation}
The meron ensembles to be considered in this study contain field
configurations obtained by superposition of  merons and antimerons of fixed
and equal number $N_M/2$,
\begin{equation}
    \label{supo}
    A_{\mu}(x)= \sum_{i=1}^{N_M} h(i) a_{\mu}(x-z(i)) h^{-1}(i).
\end{equation}
Such a configuration is specified by the position of the centers
$z(i)$ and the
color orientations
\begin{equation}
    \label{color}
    h(i) = h_0(i) + {\rm i} {\bf h}(i) \cdot \mbox{\boldmath$\sigma$} \ , \qquad
    h_0^2(i)  + {\bf h}^2(i) = 1 .
\end{equation}
In the ensembles to be discussed, the location of the merons is restricted to
a hypercube
$$   -1\le z_{\mu}(i)\le 1,\quad V=16 \, .$$
We identify the effective action in (\ref{pathintegral}) with the Yang Mills action. Our standard choice  for  meron size and coupling constant is
$$  \rho = 0.16\,, \quad g^2 =32. $$
It is essential to note that the infrared divergence of the action of a single
meron does not prevent  construction of physical ensembles  having an
extensive action with a  large number of merons. For instance, with
the following choice of the color orientations of a system of 4 merons (or 4
antimerons)
\begin{equation}
\label{quartet}
h_k(i)=\delta_{i-1,k}\, ,
\end{equation}
the action density decays as $1/x^6$.

The meron ensembles have been generated by  Monte Carlo
sampling of the action in the path integral,  Eq. (\ref{pathintegral}).
In  each step
of a Metropolis update, the position and color orientation of a given
meron are tentatively changed, the induced changes in the action density are
evaluated at a set of mesh points distributed over the whole volume, and the
configuration is accepted or rejected based on the global change in action. The
long range nature of the meron fields makes the changes extend throughout the
whole system.\\
We first present the results for the central quantity in the discussion of
confinement, the Wilson loop
\begin{equation}
    \label{wilo}
     W = \frac{1}{2} \,\mbox{tr}\left\{ P \exp {\rm i}
\oint_{\cal C}   {\rm d}x^\mu A_{\mu}(x) \right\}\, .
\end{equation}
The integral is ordered along the closed path ${\cal C}$. Our standard
choice is a rectangular path located in an $(x_i, x_j)$-plane with the center
at the origin and with the ratio of the sides equal to 2. For a given
configuration, we  evaluate twelve different Wilson loops $W_{ij}$ and obtain
our final results by taking the ensemble average and the average over the
twelve  orientations. The statistical errors are calculated from the 
variance of
the  twelve orientations.
\begin{figure}
\hspace{3cm}\epsfig{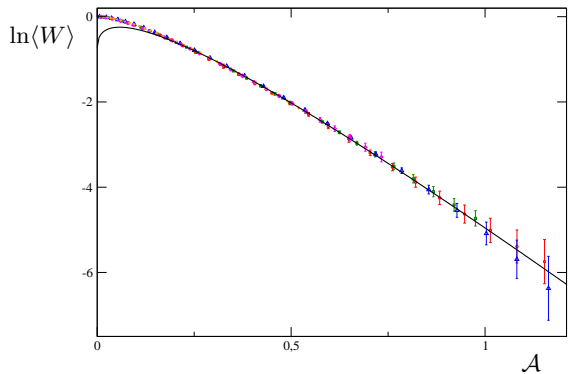}
\vskip -4.5cm \hskip -7.5cm
$\ln \langle W \rangle $
\vskip 4.cm\hskip 5.5cm ${\cal A}$
   \caption{ Logarithm of Wilson loop  as function of area ${\cal A}$.  The data
correspond to  meron ensembles at $g^2$ = 32 with  $N_M =$
1000, 500, 200,100, and 50. The area has been
rescaled by
1.86, 1.41, 1.0, 0.75, and 0.53  respectively.  The line shows the 
fit with Eq. (\ref{fit}).}
\label{wiscal}
\end{figure}
Fig. \ref{wiscal} shows the universal behavior  of Wilson loops calculated with
meron ensembles with a range of  meron numbers.
After rescaling the area $ {\cal A}\rightarrow \lambda \, {\cal A}$, 
the values of the Wilson loop lie on a universal  scaling curve, 
where deviations
from scaling
are within the statistical uncertainties. For  loops that are not too
small, the
universal curve can be parameterized by the sum of a perimeter term and an
area term
\begin{equation}
    \label{fit}
\ln  \langle W \rangle = \omega + \tau \, \sqrt{\lambda}\,{\cal P} -\sigma\,
\lambda \,{\cal A}
\end{equation}
as shown in Fig. \ref{wiscal}. The values of the parameters are
\begin{equation}
    \label{par}
\omega = -0.72,\quad \tau=0.93,\quad  \sigma = 8.17\;.
\end{equation}

The concave
shape of $\ln  \langle W \rangle$ as a function of the area arises from the
difference in sign between the area and perimeter terms  and makes the
presence of a positive string tension unambiguous.
  Numerical results
also confirm the expected  increase  of the Wilson loop with increasing
perimeter at fixed area. These  results  thus demonstrate confinement in these
meron ensembles.

\section{Topological Susceptibility and Action Density}

In view of this impressive scaling behavior, it is useful to scale 
all subsequent
results with a physical scale set by the string tension.  For convenience in
thinking about physical magnitudes,  although we are considering 
SU(2),  we will
identify the  value  of $\sigma$ with the physical value $4.2$ fm$^{-2}$ of the
string tension.
Ultimately we
will  compare dimensionless ratios involving powers of $\sigma$
with SU(2) lattice results, so that this arbitrary physical scale in 
no way affects
our quantitative results.
Thus,         our unit of
length (u. l.) is converted to physical units by
\begin{equation}
   \label{ul}
1 \ \mbox{u.\,l.} = 1.4 \,\sqrt{\lambda}\,\mbox{fm} .
\end{equation}
\begin{table}
\begin{center} 
\begin{tabular}{||c|c|c|c||}  \hline \hline
$n_{M}$  &$\rho$& $\langle  s\rangle\;\;$  & $\chi^{1/4}$ \\  \hline
[fm $^{-4}]$ &[fm$^{-1}$]&[fm $^{-4}]$  &[MeV]   
\\ \hline \hline
4.8 & $ 0.30 $ & $ 222 $ &  174\\ \hline
4.1&$ 0.26$ & $234 $ &171\\ \hline
3.3&$ 0.22$ & $237 $ & 178\\ \hline
2.9& $ 0.19$ & $ 238$ &178\\ \hline
2.9& $ 0.16 $ & $240$&  181\\ \hline \hline
\end{tabular}
\end{center}
\caption{Meron density, meron size, action density
 and topological susceptibility for meron ensembles
containing $N_M$ =1000, 500, 200, 100,  and 50 merons}
\vskip -.2cm
\label{table}\end{table} 
Table \ref{table} summarizes our principal results for the action density
$\langle s
\rangle$, and the topological susceptibility $\chi$
calculated in meron ensembles 
for different values of the meron density $n_M=N_M/V$  and scaled to the string tension. 
 A striking feature is the fact that the action density and topological susceptibility
depend essentially on a single  scale, which we  have chosen as the string
tension. 
Indeed,  despite the changes in the number of merons by up 
to  a factor
of 20, and of  meron  size and meron
density by factors of $2$ and $3$ respectively,  the action  density and
topological susceptibility vary by less than
$\pm 5\%$. The results turn out to be similarly insensitive to large changes  in the coupling constant.  
Hence, we will write the action density and topological susceptibility as
\begin{eqnarray}
    \label{scalsig}
\langle  s\rangle\ &=&13.1\;\sigma^2 \;\tilde{s}(n\rho^4,g^2) \ ,\nonumber\\
\chi^{1/4} &=& 0.44\,\sigma^{1/2}\,\tilde{\chi}^{1/4}(n\rho^4,g^2)\;.
\end{eqnarray}
where,  within five percent 
$\tilde{s}\approx 1,\; \tilde{\chi}^{1/4}\approx 1 \, . $ 
 In the regime of parameters investigated, breaking  the scale invariance
of the meron fields by introduction of the meron size  $\rho$ appears
to have  an effect  similar to that of introducing a finite lattice 
size  in the
quantum theory. In both cases, the  underlying scale invariance 
manifests itself
in the appearance  of logarithmic singularities for vanishing coordinate
space  regulators.

To understand the interplay between  meron size and   meron density we observe that the action density in the center of a  meron [cf.
Eq.(\ref{acdme})] is much larger than  the average action  density in the meron
ensembles
\begin{equation}
    \label{rtac}
\frac{s(0)}{\langle  s\rangle} = \frac{0.87}{\rho^4\sigma^2}\Bigg|_
{\rho=0.25 \,fm }= 13.5  \; .
\end{equation}
This suggests separation of the action density into  background and
meron peak contributions,
\begin{equation}
    \label{sep}
\langle  s\rangle = s_B + s_M\, .
\end{equation}
The meron peak contribution is obtained by integrating $s(x) $  
[Eq.~(\ref{acdme})] over a sphere of radius $r$,  which for small meron
size ($\rho\ll r)$ becomes
\begin{equation}
  \label{sm}  
s_{M}=n\int ^{r} {\rm d}^4 x  s(x)\rightarrow 3\pi^2 n\left(\frac{5}{12}+\ln
\frac{r}{\rho}\right) .
\end{equation}
The  matching requirement on  $r$
 \begin{equation}
  \label{rm} s(r) = s_{B}\end{equation} 
   yields the following expression for the action density
$$ \langle  s\rangle = \Big[s_B+
n\frac{3\pi^2}{4}\Big(\frac{5}{3}-\ln \frac{2}{3}  s_B \rho^4
\Big)\Big].$$
This expression makes explicit the logarithmic singularity of  $\langle  s\rangle$ with
the meron size in the small $\rho$ limit if $s_B$ is identified with the physical (i.e. regularized)  value  of the action density. 
With  the action density $s$ as an input (cf. Table \ref{table}), the relative strength of the background contribution can be determined numerically  
$$\frac{s_B}{\langle  s\rangle} = 0.65 \, -\, 0.75\, .$$
This result indicates that a significant fraction of the action density is associated with
 the logarithmically singular contribution in Eq. (\ref{sm}).

We can now compare the meron results in Eq.(\ref{scalsig}) with  known QCD results. The topological susceptibility 
\begin{equation}
\label{susc}
\chi = \Big(\frac{1}{32\pi^2}\Big)^2 \int {\rm d}^4x \, \langle\, F^a_{\mu\nu}\tilde{F}^a_{\mu\nu}(x)\;F^a_{\mu\nu}\tilde{F}^a_{\mu\nu}(0)\,\rangle .
\end{equation}
is a robust quantity that can be measured unambiguously in our effective theory and in lattice QCD.  It has a direct physical interpretation because of its relation to the $\eta'$ mass by the
Veneziano-Witten formula.  The most extensive lattice measurement  of the  $SU(2)$ topological susceptibility by extracting the continuum limit from calculations over a large range of the  coupling constant \cite{Lucini:2001ej} yields  $\chi^{1/4} / \sigma^{1/2}  \sim 0.483 \pm .006 $.  The meron result $\chi^{1/4} / \sigma^{1/2}  \sim 0.44 $ from  Eq.(\ref{scalsig}) is in excellent agreement with this lattice result.
Note that $\chi$ can be computed reliably for the meron ensembles since the topological charge of a single meron, Eq. (\ref{topc}),  is  finite. 
The topological susceptibility is  dominated by the
short-range peaks in the topological charge density associated with individual
merons and antimerons, and not by the long range background field. The contribution due to single peaks yields
\begin{equation}
\label{susc2}
   \chi^{1/4}= 0.505 \,n_M^{1/4} .
\end{equation}
The weak variation in $ \chi^{1/4}$ and the magnitude of this estimate are in  qualitative agreement
  with the results of Table \ref{table}. 

Because of the necessity of subtracting divergent terms to define the continuum limit, the action density, or equivalently the gluon condensate, is more difficult than the topological susceptibility to evaluate accurately in either QCD or a meron ensemble. Lattice   $SU(2)$ calculations  for the action density range from $\langle s\rangle /\sigma^2 = 4.5$  \cite{Campostrini:1983nr} to 25.3 \cite{DiGiacomo:1989id} and QCD sum rule results range from $\langle s\rangle /\sigma^2  = 4.5 $ \cite{Shifman:bx} to 10 \cite{Narison:1995tw}. The meron result $\langle s\rangle /\sigma^2 \sim 13.1$  from  Eq. (\ref{scalsig}) is thus consistent with our present knowledge of the action density. It is useful to note that in the lattice calculations of Ref. \cite{Campostrini:1983nr},
the divergent contributions are about a factor of 20-500 larger
 than the extracted value of $\langle s\rangle$. As noted above,  the meron action density also contains 
 a divergent contribution  that,  if subtracted,  will significantly  reduce the value of the condensate given in Table \ref{table}. 
\section{Conclusions}
 In conclusion, we have shown that an effective theory with meron degrees of freedom succeeds
in describing essential features of QCD.  The long-range gauge fields
provide the correlations and disorder needed for confinement while the short
range fluctuations play an essential role in other physical observables.
They render the meron dynamics more
complicated than for ensembles of weakly interacting singular gauge
instantons and must therefore be treated numerically. 
The central feature and a major success of this approach is producing confinement.  We have 
demonstrated a confining area law and discussed how a meron ensemble can
implement center symmetry.
We also observe scaling behavior, reminiscent of lattice Yang Mills theory which might not {\it a priori}  have been
expected.  In our effective theory there are three
parameters, $g^2$, $\rho$, and $n_M$, and one combination  of  $\rho$ and
$n_M$ is determined by fitting to the string tension at a given $g^2$.  The action
density and topological charge density are rather insensitive to the other
combination and  the Wilson loops scale to a universal curve.  This scaling
behavior is presumably  connected with the scale invariance of the multi-meron
action in the limit $\rho  \rightarrow 0$.
In addition, we have shown nearly quantitative agreement with
the value of the topological susceptibility as measured in SU(2) lattice QCD and have obtained a value of the action density of the correct order of magnitude. 

 It is a pleasure to acknowledge useful conversations with Mikhail Shifman,
Edward Shuryak, and Frank Wilczek.  J.N. is grateful for support by 
an Alexander
von Humboldt Foundation Research Award and for  hospitality at the Institute
for Theoretical Physics III at the University of Erlangen where this research
was initiated.
This work was supported in part by
funds provided by the U.S. Department of Energy (D.O.E.) under cooperative
research agreement DE-FC02-94ER40818.

\end{document}